\documentclass[prd, preprintnumbers, nofootinbib, notitlepage,  preprintnumbers]{revtex4}
\pdfoutput=1
\usepackage{graphicx}
\usepackage{color}
\usepackage{mathrsfs}
\usepackage{amsmath}
\usepackage{hyperref}
\newcommand{\pprime}{{\prime\prime}}

\begin{document}
\title{More Exact Tunneling Solutions in Scalar Field Theory}
\author{Koushik Dutta}
\author{Cecelie Hector}
\author{Pascal M.~Vaudrevange}
\author{Alexander Westphal}
\affiliation{DESY, Notkestrasse 85, 22607 Hamburg, Germany}
\preprint{DESY 11-167}
\begin{abstract}
  We present exact bounce solutions and amplitudes for tunneling in i)
  a piecewise linear-quartic potential and ii) a piecewise
  quartic-quartic potential, ignoring the effects of
    gravitation. We cross check their correctness by comparing with
  results obtained through the thin-wall approximation and with a
  piecewise linear-linear potential. We briefly comment on
  applications in cosmology.
\end{abstract}
\date{\today}
\maketitle
\section{Introduction}
In recent times, first order phase transitions have gained significant
interest, for example as sources of gravitational waves
\cite{Huber:2008hg} and in transversing the string theory landscape
\cite{Bousso:2000xa}, \cite{Susskind:2003kw}. In the latter picture,
the scalar field potential possesses a plethora of local minima. A
field that is initially trapped in a higher energy vacuum jumps to a
lower energy vacuum via a quantum tunneling process.

The underlying microphysics of tunneling can be described by
instantons, i.e. classical solutions of the Euclidean equations of
motion of the system \cite{Coleman:1977py},
\cite{Coleman:1980aw}. Tunneling proceeds via the nucleation of
bubbles of true (or rather lower energy) vacuum surrounded by the sea
of false vacuum. If the curvature of the potential is large compared
to the corresponding Hubble scale, this process can be described by
Coleman de Luccia (CdL) instantons, i.e. bounce solutions to the
Euclidean equations of motion \cite{Coleman:1977py},
\cite{Coleman:1980aw}. For relatively flat potentials, tunneling
proceeds via Hawking-Moss instantons
\cite{Hawking:1981fz}. 

Ignoring the effects of gravity, Coleman presented
  a straightforward prescription for computing vacuum transitions~\cite{Coleman:1977py}.
The tunneling amplitude for a transition from the false (or higher
energy) vacuum at $\phi_+$ to the true (or lower energy) vacuum at
$\phi_-$ is given by $A\exp(-B)$. The coefficient $A$ is typically
ignored but in principle calculable, see \cite{Callan:1977pt}. The
exponent $B=S_E(\phi_B)-S_E(\phi_+)$ (sometimes also referred to as
the bounce action) is the difference between the Euclidean action
$S(\phi)=2\pi^2\int_0^\infty\!dr\,r^2
\left(\frac{1}{2}\phi^{\prime2}+V(\phi)\right)$ for the spherically
symmetric bounce solution $\phi_B$ and for the false vacuum
$\phi_+$. The bounce obeys the one-dimensional Euclidean equation of
motion
\begin{eqnarray} \label{bounce_equation}
  \phi_B^\pprime+\frac{3}{r}\phi_B^\prime-\partial_\phi V(\phi_B)&=&0\,,
\end{eqnarray}
where $\phi^\prime\equiv\partial_r\phi$ and $r=\sqrt{t^2-\vec{x}^2}$
is the radial coordinate of the spherical bubble. This configuration
describes the bubble at the time of nucleation. In this paper, we
ignore its subsequent evolution, and focus on the computation of $B$.

In general, the CdL bounce solutions can be computed exactly only for
very few potentials. However, if the potential difference between the
two vacua is small compared to the typical potential scale, the
tunneling amplitude can be computed using the thin wall
approximation. Otherwise, one needs to resort to either numerical
computations (see \cite{Adams:1993zs} for an approach for a generic
quartic potential) or approximate the potential by potentials for
which the exact instanton solutions are known. To the best of our
knowledge, only for very few potentials has the CdL tunneling process
been solved analytically: a piecewise linear-linear potential
\cite{Duncan:1992ai} and piecewise linear-quadratic potentials
\cite{Hamazaki:1995dy}, \cite{Pastras:2011zr},
\cite{Dutta:2011ej}. While the paper was being finished, we became
aware of \cite{Lee:1985uv} who presented a bounce solution for
tunneling in a quartic-linear potential. A different approach was
taken by \cite{Dong:2011gx} who reconstruct fully analytically
tractable potentials, including the effects of gravity, from
analytically exact bubble geometries.

We present new exact solutions for tunneling within piecewise
potentials where the true vacuum potential is a quartic, see Figures
\ref{fig:linear_quartic:potential} and
\ref{fig:quartic_quartic:potential}. The potential for $\phi>0$ (``on
the right'') is given by
\begin{eqnarray}
  V_R (\phi)&=&V_T - \Delta V_- +\frac{\Delta V_-}{\phi_-^4}\left(\phi-\phi_-\right)^4\,,
\end{eqnarray}
where $\Delta V_-\equiv V_T-V_-$. For simplicity, we chose $\phi=0$ as
the matching point and $V(\phi = 0) = V_T$. We will choose the
potential for $\phi<0$ (``on the left'') as either linear or quartic
and discuss the solutions in Section \ref{sec:linear_quartic} and
Section \ref{sec:quartic_quartic} respectively.

For each piecewise potential, we proceed analogously to
\cite{Duncan:1992ai}, \cite{Dutta:2011ej}: First we solve the equation
of motion for the scalar field in $V_R (\phi)$, subject to the
boundary condition at the center of the bubble $\phi_R(0)=\phi_0,
\phi_R^\prime(0)=0$. We assume that the bubble nucleation point is
located at $\phi_0>0$, i.e. it is in the valley of the true
vacuum. Then, we solve the equation of motion for the field in $V_L$,
subject to $\phi_L(R_+)=\phi_+, \phi_L^\prime(R_+)=0$. In other words,
we assume that at some radius $R_+$ (which can be $\infty$) outside of
the bubble of true vacuum, the field sits in the false vacuum. Then,
we match the solutions at some radius $R_T$ by enforcing
$\phi_L(R_T)=\phi_R(R_T)=0$ and
$\phi_L^\prime(R_T)=\phi_R^\prime(R_T)$. This allows us to determine
the constants $R_T, R_+$, and $\phi_0$. Here, $R_T$ is roughly the radius of
the bubble when it materializes at $\phi =\phi_0$, whereas the value
comparing $R_{+}$ to $R_T$ gives us an idea about the width of the
bubble wall. 

It is then straightforward to integrate the action for $\phi_L$ and
$\phi_R$, obtaining $B$. We compare the tunneling bounce action $B$ for
the piecewise linear-quartic potential potential with the results of
both the thin-wall approximation and the piecewise linear-linear
potential solved in \cite{Duncan:1992ai}. Finally, we compute the
tunneling amplitude for the piecewise quartic-quartic potential and
compare it with the results obtained using the thin-wall
approximation, as well as with the tunneling amplitude in a piecewise
linear-quartic potential.

\section{Linear on the left, quartic on the right}\label{sec:linear_quartic}
\begin{figure}
  a)\includegraphics[width=0.45\textwidth]{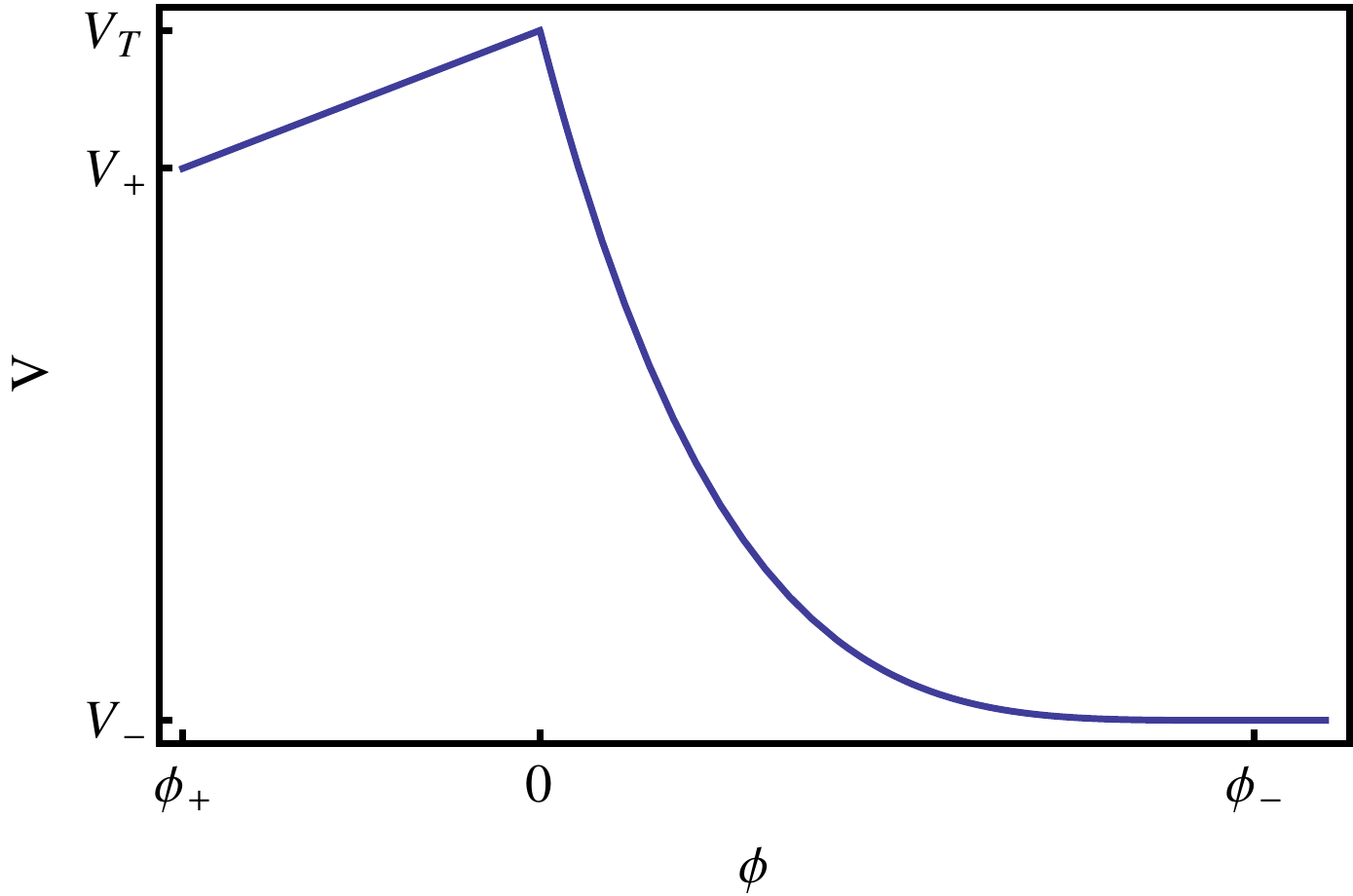}
  b)\includegraphics[width=0.45\textwidth]{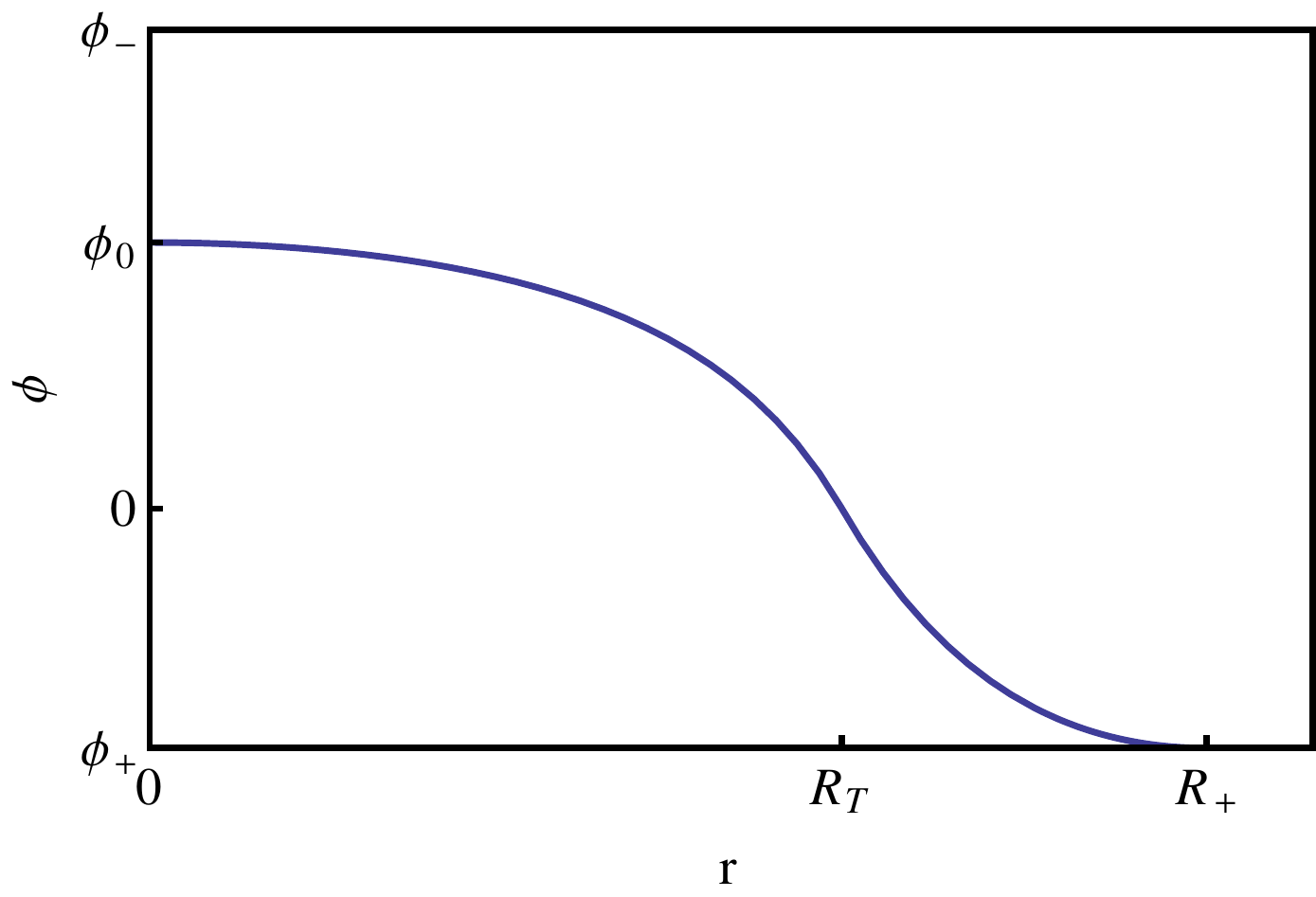}
  \caption{a) Schematic plot of the piecewise linear-quartic
    potential. The left part of the potential is a linear function of
    $\phi$, the right part a quartic function. The bounce describes
    tunneling from the field sitting in the false vacuum at $\phi_+$
    towards the true vacuum located at $\phi_-$. b) Schematic view of
    the bounce solution for (a). Inside the bubble at $r=0$, the field
    is at $\phi_0>0$. The bubble wall is located around $R_T$, but not
    necessarily thin. Outside of the bubble at $r=R_+$, the field is
    still in the false vacuum.}
  \label{fig:linear_quartic:potential}
  \label{fig:linear_quartic:bounce}
\end{figure}

In this section we compute the tunneling rate for a piecewise
potential of the form
\begin{eqnarray}
  V(\phi)&=&\left\{\begin{array}{ll}
  V_T - \frac{\Delta V_+}{\phi_+}\phi\,,& \phi\le0\,,\\
  V_T - \Delta V_- +\frac{\Delta V_-}{\phi_-^4}\left(\phi-\phi_-\right)^4\,&\phi>0\,,
  \end{array}
  \right.
\end{eqnarray}
where $\Delta V_- \equiv V_T-V_-=\frac{\lambda_4}{4} \phi_-^4$ and
$\Delta V_+\equiv V_T-V_+=-\lambda_1\phi_+$ are the depths of the true
and false minimum, see Figure~\ref{fig:linear_quartic:potential}.
Subject to the boundary conditions $\phi_R(0)=\phi_0,
\phi_R^\prime(0)=0$, solving the equation of motion of the bounce,
i.e. Eq.~\eqref{bounce_equation} on the right side of the potential,
we have \cite{Dutta:2011fe}
\begin{eqnarray}\label{eq:phi_quartic_generic}
  \phi_R(r)&=&\phi_-+\frac{2(\phi_0-\phi_-)}{2-\frac{ \Delta V_-(\phi_0-\phi_-)^2}{\phi_-^4}r^2}\,.
\end{eqnarray}
Similarly on the left side of the potential,
subject to $\phi_L(R_+)=\phi_+, \phi_L^\prime(R_+)=0$, we have the bounce solution
\begin{eqnarray}
  \phi_L(r)&=&\phi_+-\frac{\Delta V_+}{8\phi_+}\frac{(r^2-R_+^2)^2}{r^2}\,.
\end{eqnarray}
A schematic view of the bounce is shown in Figure~\ref{fig:linear_quartic:bounce} b).

We now determine the constants $R_+$ and $\phi_0$ by solving the matching
equations for the two solutions $\phi_R(R_T)=0,\,\phi_L(R_T)=0$.
Using the first condition, we get $\phi_0$ in terms of $R_T$
\begin{eqnarray}\label{eq:phi0}
  \phi_0&=&\frac{\phi_-^3}{\Delta V_- R_T^2}\left[\frac{\Delta V_- R_T^2}{\phi_-^2}+\left(1-\sqrt{\frac{2\Delta V_- R_T^2}{\phi_-^2} + 1}\right)\right]\,,
\end{eqnarray}
while the second condition gives
\begin{eqnarray}
  R_+&=&\sqrt{R_T\left(R_T+\frac{2\sqrt{2}\alpha\phi_-}{\sqrt{\Delta \Delta V_-}}\right)}\,.
\end{eqnarray}
Here, we have introduced $\Delta=\Delta V_+/\Delta V_-$ and
$\alpha=-\phi_+/\phi_-$.  Similarly, using the smoothness of the
solution at $R_T$, i.e. $\phi_R^\prime(R_T) = \phi_L^\prime(R_T)$, we
find
\begin{equation}\label{eq:Rt}
  R_T=\frac{\phi_- \left(\sqrt{\Delta}(1+2\alpha)+\sqrt{4\alpha(1+\alpha)+\Delta}\right)}{(1-\Delta)\sqrt{2\Delta V_-}}\,.
\end{equation}
Computing the exponent of the tunneling amplitude in terms of $R_T$
gives
\begin{widetext}
\begin{eqnarray}
  B&=&\frac{\pi^2}{6\Delta V_-}\left\{3R_T^4\left(\Delta-1\right)\Delta V_-^2+8\sqrt{2}R_T^3\alpha\Delta V_-\sqrt{\Delta\Delta V_-}\phi_-+2\phi_-^4\left[-1+\sqrt{1+\frac{2R_T^2\Delta V_-}{\phi_-^2}}\right]\right.\nonumber\\
  &&\phantom{bla}\left.+2R_T^2\Delta V_- \phi_-^2\left[(6\alpha^2-3)+2\sqrt{1+\frac{2R_T^2\Delta V_-}{\phi_-^2}}\right]\right\}\label{eq:BinRt}\,.
\end{eqnarray}
Plugging $R_T$ from Eq.~\eqref{eq:Rt}, we obtain a rather monstrous
expression
\begin{eqnarray}\label{eq:linear_quartic:B:full}
  B&=&\frac{\pi^2\phi_-^4}{6\Delta V_-}\Bigg\{4\alpha\sqrt{\Delta}\left[\frac{(1+2\alpha)\sqrt{\Delta}+\sqrt{4\alpha(1+\alpha)+\Delta}}{1-\Delta}\right]^3-\frac{3}{4}\left[\frac{(1+2\alpha)\sqrt{\Delta}+\sqrt{4\alpha(1+\alpha)+\Delta}}{(1-\Delta)^{3/4}}\right]^4\nonumber\\
  &&\phantom{bla}+\left[\frac{(1+2\alpha)\sqrt{\Delta}+\sqrt{4\alpha(1+\alpha)+\Delta}}{1-\Delta}\right]^2\left[-3+6\alpha^2+2\sqrt{1+\left[\frac{(1+2\alpha)\sqrt{\Delta}+\sqrt{4\alpha(1+\alpha)+\Delta}}{1-\Delta}\right]^2}\right]\nonumber\\
  &&\phantom{bla}+2\left[-1+\sqrt{1+\left[\frac{(1+2\alpha)\sqrt{\Delta}+\sqrt{4\alpha(1+\alpha)+\Delta}}{1-\Delta}\right]^2}\right]\Bigg\}\,.
\end{eqnarray}
\end{widetext}

To cross check our result, we take the thin-wall limit of
Eq.~\eqref{eq:linear_quartic:B:full} by replacing
$\Delta=1-\frac{\epsilon}{\Delta V_-}$, where $\epsilon$ is the energy
difference between the true and false vacua. In the thin-wall limit
$\epsilon \ll V_T$. Performing a series expansion around $\epsilon=0$,
the lowest order term in $\epsilon$ is
\begin{equation}\label{eq:linear_quartic:B:thin_wall}
  \lim_{\epsilon\to0}B=\frac{2\pi^2}{3}\frac{(1+2\alpha)^4\phi_-^4\Delta V_-^2}{\epsilon^3}\,.
\end{equation}
We compare this with the results obtained using the thin wall
approximation \cite{Coleman:1977py}
\begin{eqnarray}
  B_{TW}&\equiv&\frac{27\pi^2}{2}\frac{S_1^4}{\epsilon^3}\,,
\end{eqnarray}
where
\begin{eqnarray}
  S_1&\equiv&\int_{\phi_-}^{\phi_+}\!d\phi\sqrt{2\left(V(\phi)-V(\phi_+)\right)}=-\frac{\sqrt{2\Delta V_-}\phi_-}{3}\left[(1+2\alpha)\sqrt{\Delta}+2\sqrt{\Delta-1}\,_2F_1\left(\frac14,\frac12,\frac54,\frac{1}{1-\Delta}\right)\right]\,,
\end{eqnarray}
with hypergeometric function $_2F_1$. Again, replacing
$\Delta=1-\frac{\epsilon}{\Delta V_-}$ gives to the lowest order in
$\epsilon$
\begin{eqnarray}\label{Eq:linear_quartic_TH}
  B_{TW}&\approx&\frac{2\pi^2}{3}\frac{(1+2\alpha)^4\phi_-^4\Delta V_-^2}{\epsilon^3}\,,
\end{eqnarray}
in agreement with Eq.~\eqref{eq:linear_quartic:B:thin_wall}. 

As another cross-check\footnote{Comparing our results with
  \cite{Lee:1985uv}, we find that the tunneling rate is quite
  different. This can be traced back to the fact that tunneling from a
  quartic into a linear potential should reduce to the $\alpha<1$
  solution of Duncan and Jensen in the appropriate limit.}, we observe
that for fixed $\Delta$ and $\phi_+$, sending $|\phi_-|\ll|\phi_+|$,
the potential on the right appears more and more like a linear
potential. In other words, in the limit of $\alpha\gg1$, the tunneling
bounce action in Eq.~\eqref{eq:linear_quartic:B:full} must agree with the
tunneling bounce action in a piecewise linear-linear potential.  The exact
tunneling amplitude for a piecewise linear-linear potential has been
calculated by Duncan and Jensen \cite{Duncan:1992ai}. In our notation,
their result for $\alpha>1$ is given by
\begin{eqnarray}
 B_{DJ}&=&\frac{2\pi^2}{3}\left(\frac{1+\alpha}{\sqrt{\Delta}-1}\right)^3\frac{\phi_-^4}{\Delta V_-}\left[(\alpha-3)\sqrt{\Delta}+1-3\alpha\right]\,.
\end{eqnarray}
In the limit of large $\alpha\gg1$, i.e. for $|\phi_-|\ll|\phi_+|$,
this becomes
\begin{eqnarray}
  \lim_{\alpha\to\infty}B_{DJ}&=&\frac{2\pi^2}{3}\frac{\alpha^4(\sqrt{\Delta}-3)}{(\sqrt{\Delta}-1)^3}\frac{\phi_-^4}{\Delta V_-}\,, 
\end{eqnarray}
which indeed agrees with the corresponding limit of
Eq.~\eqref{eq:linear_quartic:B:full}. Note that this is
independent of the thin-wall limit.

As an aside, we observe some curious systematic behavior: the radius
of the bubble in the thin-wall limit for a piecewise linear-quartic
potential is given by
\begin{equation}
R_T= \frac{3S_1}{\epsilon} = (1+2\alpha)\frac{\sqrt{2\Delta V_-}\phi_-}{\epsilon}\,.
\end{equation}
For a cubic potential for $V_R(\phi)$ on the right, the thin-wall
approximation gives
\begin{eqnarray}
  R_T&=&\left(\frac{6}{5}+2\alpha \right)\frac{\sqrt{2\Delta V_-}\phi_-}{\epsilon}\,.
\end{eqnarray}
Finally, for $V_R(\phi)$ a quadratic potential, the bubble radius is
given by \cite{Dutta:2011ej}
\begin{equation}
R_T = \left(\frac{3}{2}+2\alpha\right)\frac{\sqrt{2\Delta V_-}\phi_-}{\epsilon}.
\end{equation}
Thus we find that in the thin wall approximation, the nucleated bubble
size shrinks mildly as the power of the monomials for potential in the
exiting part (near the true vacuum) becomes larger.

\section{Quartic on the left and quartic on the right}\label{sec:quartic_quartic}

\begin{figure}
  a)\includegraphics[width=0.45\textwidth]{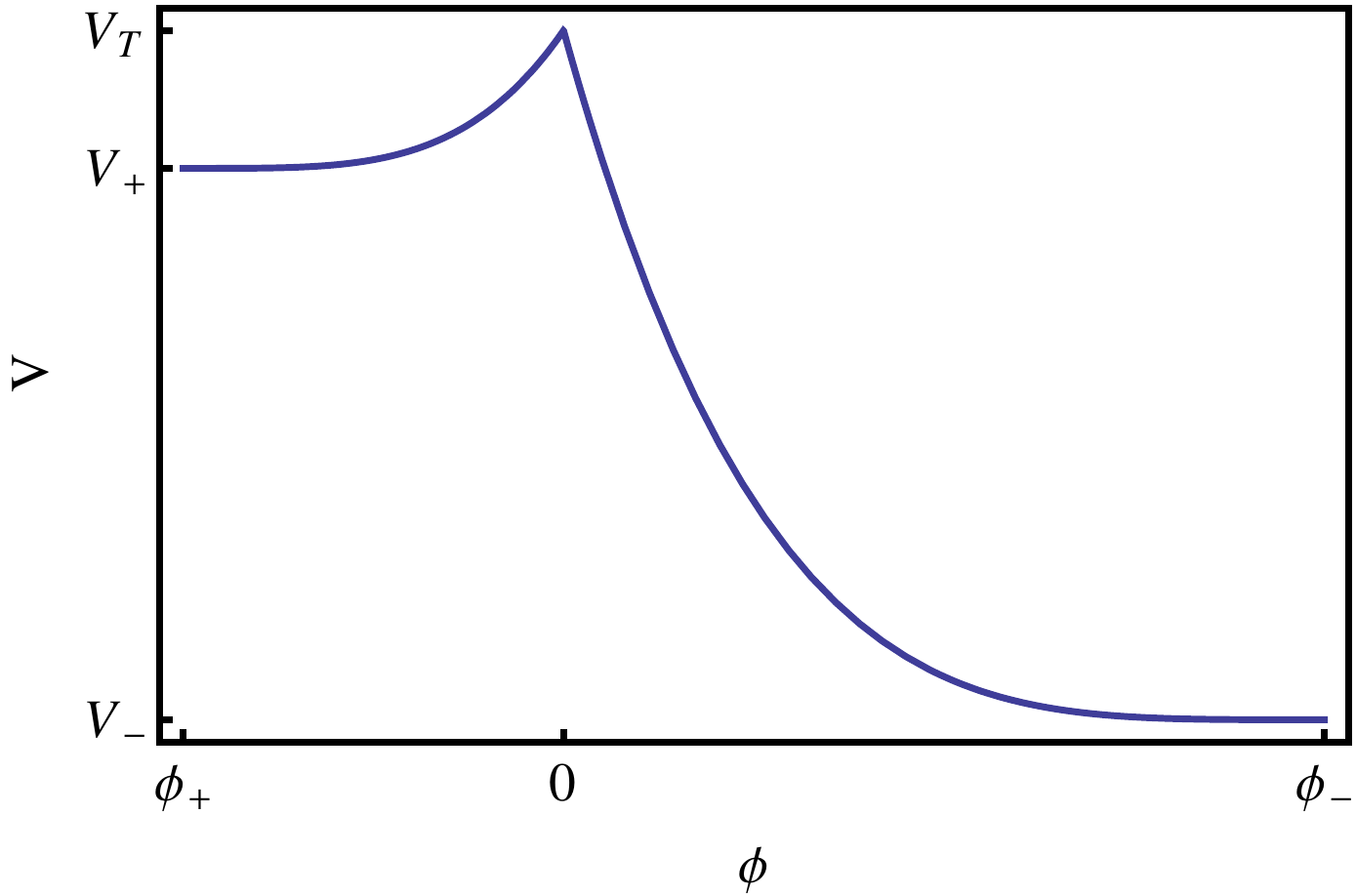}
  b)\includegraphics[width=0.45\textwidth]{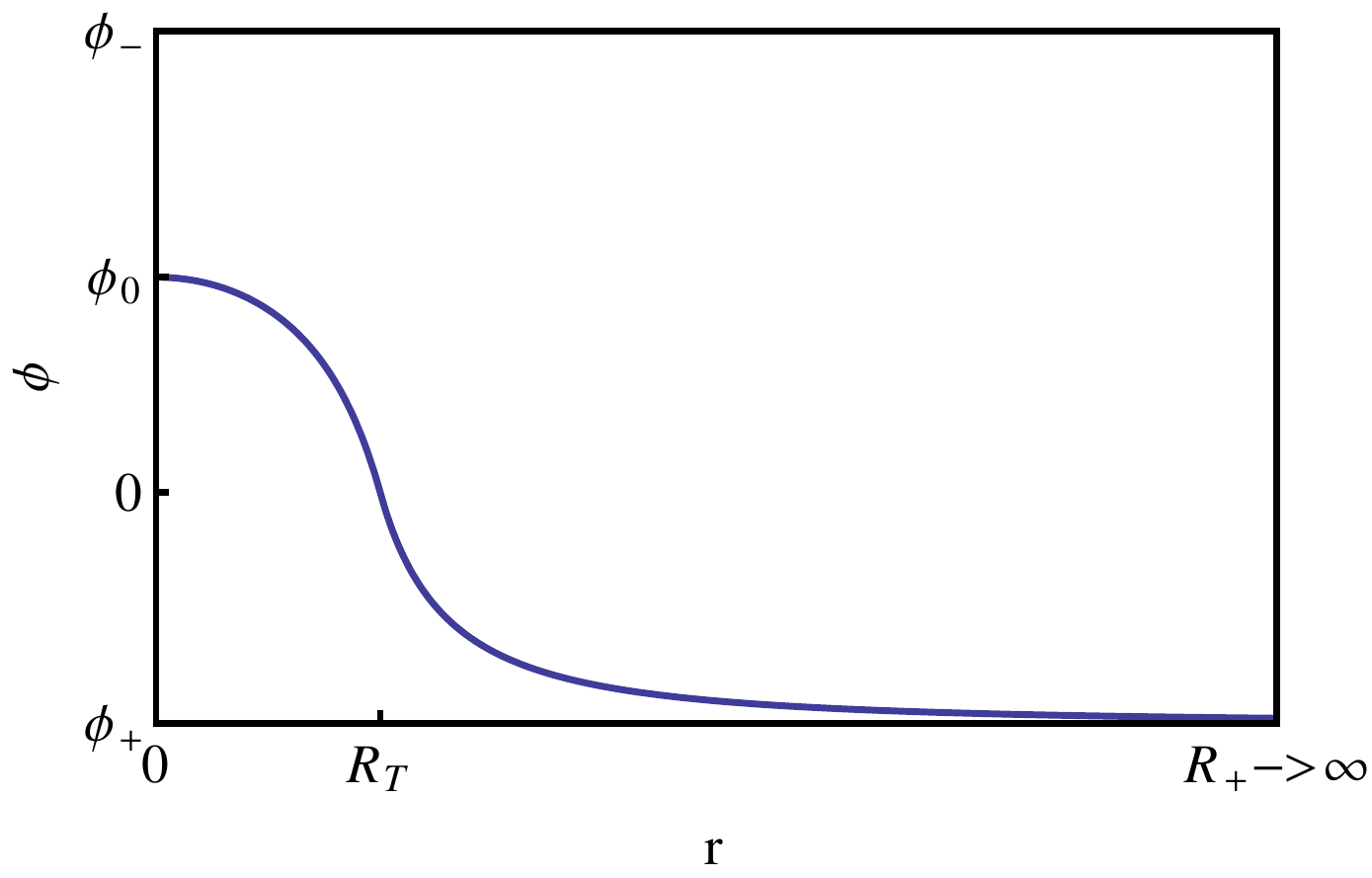}
  \caption{a) Schematic view of a piecewise quartic-quartic
    potential.  b) Schematic view of the bounce solution for a). Note
    that the position where the field is still in the false vacuum
    goes to infinity, $R_+\rightarrow\infty$.}
  \label{fig:quartic_quartic:potential}
\end{figure}

In this section, we compute the bounce solution for tunneling from the
false vacuum in a quartic potential to the true minimum in another
quartic potential, see Figure~\ref{fig:quartic_quartic:potential}a).
We can reuse parts of the previous calculation, in particular the
solution inside the bubble from Eq.~\eqref{eq:phi_quartic_generic} and
Eq.~\eqref{eq:phi0}. Outside of the bubble, the field sits in the
false vacuum
\begin{eqnarray}
  \phi_L(R_+)&=&\phi_+\quad \phi_L^\prime(R_+)=0\,.
\end{eqnarray}
Note that, if we are not interested in knowing the width of the
bubble, the boundary conditions above can also be set at $r \to
\infty$. It turns out that this is what we need to do. The solution
$\phi_L(r)$ has the form
\begin{eqnarray}
  \phi_L(r)&=&\phi_++\frac{8A}{8-\frac{4 \Delta\Delta V_- A^2 r^2}{\phi_+^4}}\,,
\end{eqnarray}
with $A$ to be fixed by the condition that $\phi_L(R_T)= 0$. Thus we
find
\begin{eqnarray}
  \phi_L(r)&=&\frac{\left(r^2-R_T^2\right)\alpha\phi_-\left(\frac{\Delta R_T^2 \Delta V_-}{\alpha^2\phi_-^2}+\left(1+\sqrt{\frac{2\Delta R_T^2\Delta V_-}{\alpha^2\phi_-^2}+1}\right)\right)}{\Delta R_T^2\alpha^2\frac{\Delta R_T^2 \Delta V_-}{\alpha^2\phi_-^2}-r^2\left(\frac{\Delta R_T^2\Delta V_-}{\alpha^2\phi_-^2}+\left(1+\sqrt{\frac{2\Delta R_T^2\Delta V_-}{\alpha^2\phi_-^2}+1}\right)\right)}\,.
\end{eqnarray}
From the smoothness of the solution $\phi_L^\prime(R_T) =
\phi_R^\prime(R_T)\,$ we obtain
\begin{eqnarray}
  R_T&=&\frac{\sqrt{2(1+\alpha)(\alpha+\Delta)}}{1-\Delta}\frac{\phi_-}{\sqrt{V_-}}\,.
\end{eqnarray}

Integrating the Euclidean action gives
\begin{eqnarray}\label{eq:quartic_quartic:B:full}
  B&=&\frac{2\pi^2}{3}\frac{4\alpha^3+6\alpha^2\Delta+4\alpha\Delta^2+\Delta^3+\alpha^4(3+\Delta(\Delta-3))}{(1-\Delta)^3}\frac{\phi_-^4}{\Delta V_-} \,,
\end{eqnarray}
which in the thin-wall limit reduces to
\begin{eqnarray}\label{eq:quartic_quartic:B:thin_wall}
  B&\approx&\frac{2\pi^2}{3}\frac{(1+\alpha)^4\Delta V_-^2\phi_-^4}{\epsilon^3}\,.
\end{eqnarray}
Using the thin wall formula we find
\begin{eqnarray}
  S_1&=&-\frac{\sqrt{2\Delta V_-}}{3}\left[(1+\alpha)\sqrt{\Delta}+2\sqrt{(\Delta-1)}\,_2F_1\left(\frac14,\frac12,\frac54,\frac{1}{1-\Delta}\right)\right]\,,
\end{eqnarray}
and in the small $\epsilon$ limit $B$ agrees with
Eq.~\eqref{eq:quartic_quartic:B:thin_wall}.

We note that in the thin-wall limit, the tunneling bounce action $B$ for
tunneling in a piecewise linear-quartic potential differs from the
one in a piecewise quartic-quartic potential by the substitution
$\alpha\to2\alpha$. In particular, this means that for $\alpha\gg1$,
tunneling in a piecewise linear-quartic potential is much more
suppressed than tunneling in a piecewise quartic-quartic
potential: the respective values of $B$ differ by a factor of $16$,
suppressing the relative amplitude by the $16^\mathrm{th}$ power.

\begin{table}
  a)\quad\begin{tabular}{l|l|l|l}
    &$B_{\text{ll}}$&$B_{\text{lq}}$&$B_{\text{qq}}$\\
    \hline
    $\alpha=0.01$&0.0072&0.00024&0.00010\\
    $\alpha=0.1$&0.4&0.058&0.033\\
    $\alpha=0.5$&24&6.7&4.8
  \end{tabular}
  \qquad b)\quad\begin{tabular}{l|l|l}
    &$B_{\text{qq}}$&$B_{\text{qq, thin-wall}}$\\
    \hline
    $\Delta=0.99$&$3.3\times10^7$ & $3.3\times10^7$\\
    $\Delta=0.9$&$2.8\times10^4$ & $3.3\times10^4$\\
    $\Delta=0.7$&$7.2\times10^2$ & $1.2\times10^3$
  \end{tabular}
  \caption{a) Tunneling bounce actions for different values of
      $\alpha$ with $\Delta=0.01$ fixed. Tunneling in a linear-linear
      potential is consistently suppressed compared to tunneling in
      linear-quartic and quartic-quartic potentials -- keeping in mind
      that a larger $B$ corresponds to smaller tunneling rates. b)
      Comparison with the thin-wall approximation for tunneling in a
      quartic-quartic potential for fixed $\alpha=0.5$. Decreasing
      $\Delta$ away from unity (i.e. exact equality between false and
      true vacuum energy), it is clear that the thin-wall
      approximation eventually fails. }
  \label{tab:tunneling_amplitudes}
\end{table}
To further explore the differences in tunneling rates for
  different potential shapes, we tabulate the values for $B$ for
  different values of $\alpha$, keeping $\Delta=0.01$ fixed for
  tunneling in a linear-linear (ll), linear-quartic (lq), and
  quartic-quartic (qq) potential, see
  Table~\ref{tab:tunneling_amplitudes}. For all values of $\alpha$,
  the width of the wall of the nucleated bubble is non-negligible,
  $(R_+-R_T)/R_T=O(1)$, so we are dealing with tunneling in the thick-wall regime. As can be seen, the action $B$ for tunneling in a
  linear-linear potential are always significantly larger than for
  tunneling in linear-quartic and quartic-quartic potentials. As the
  tunneling rate is proportional to $e^{-B}$, even ${\cal O}(1)$ factors lead to significant differences of
  the tunneling rates. In the thick-wall regime tunneling seems to depend crucially
  on the exact shape of the potential, making the search for more exact
  tunneling solutions even more pressing.

\section{Conclusions}
In this brief article, we discuss a quantum tunneling event in a
piecewise potential where the false vacuum part is either linear or
quartic and the true vacuum is described by a quartic
potential. Often, the analysis of quantum tunneling in field theory is
performed in the thin wall approximation \cite{Coleman:1977py}. This
does not necessarily capture all realistic scenarios. In particular,
cosmological phase transitions usually involve a large change of the
energy scale. For example, the relative energy difference between
neighboring vacua in the landscape of string theory is typically
large. Although any specific realistic scenario can be solved by
numerical methods, this makes it rather difficult to get a good
qualitative understanding of the process under a change of potential
parameters. As shown in the previous section, the exact shape of
  the potential plays a non-negligible role when considering tunneling
  in the thick-wall regime.  Together with previous exact
tunneling solutions \cite{Duncan:1992ai}, \cite{Hamazaki:1995dy},
\cite{Pastras:2011zr}, \cite{Dutta:2011ej}, this work contributes to
bridging the gap in qualitative understanding. As a consistency check,
we have shown that the tunneling rates always reduce to the thin-wall
result in the appropriate limit.

It may be appropriate at this point to outline, that our exact results here for tunneling in a piecewise linear-quartic or quartic-quartic potential can be used to describe analytically models of open inflation in a toy landscape constructed from piecewise linear and quartic potentials. The toy inflationary landscape is constructed from a piecewise linear-quartic or quartic-quartic potential, to which a slow-roll inflationary region is attached with matching $V'$ at $\phi\simeq\phi_-$. The crucial point here is that the quartic potential which dominates field evolution after tunneling and before entering the slow-roll region, completely suppresses a would-be fast-roll overshoot problem in the slow-roll region. This happens because the negative spatial curvature inside the CdL bubble (once gravity is to be included~\cite{Coleman:1980aw}, which we -- but for the negative curvature inside the bubble -- do not discuss here) formed during tunneling provides a very strong friction term. This Hubble friction is sufficient for damping the downhill motion enough to start slow-roll subsequently~\cite{Dutta:2011fe} for any potential
\begin{equation}
V(\phi)=V_0+(\phi-\phi_-)^p\quad,\quad p\geq 4\quad.
\end{equation}
In such potentials the field will reach slow-roll already at some $\phi<\phi_-$ without overshoot, if the field starts its evolution inside a negatively curved CdL bubble following tunneling. Because of this fact, it does not matter whether the slow-roll inflationary region in the scalar potential at $\phi\gtrsim\phi_-$  will describe a small-field or large-field inflation model, as all models are treated equally in this toy landscape. We can now take a look at the situation where the barrier parameters $\alpha,\Delta$ take values in a dense discretuum specified in terms of a dense discretuum of microscopic parameters of a landscape of isolated vacua, such as the landscape of string theory vacua. For the moment, we will keep $\alpha$ fixed, as at $\alpha=0$ the scalar potential becomes discontinuous and the bounce ceases to exist. We may now assign $\Delta$, which controls the aspect of the barrier shape crossing over between the thin-wall and thick-wall limit, a prior probability distribution $p(\Delta)$. This distribution contains the unknown microscopic landscape data. As explained before, all values of $\Delta$ are treated equally when it comes to the slow-roll inflationary regime attached at $\phi\gtrsim \phi_-$ in our toy landscape. Therefore, the expectation value of $\Delta$ is given by
\begin{equation}
\langle\Delta\rangle = \frac{\int d\Delta\,p(\Delta)\, e^{-B(\Delta)}}{\int d\Delta\, p(\Delta)}\quad.
\end{equation}
This does not depend on the post-tunneling inflationary dynamics due to the absence of overshoot. Therefore, in such a toy landscape the question whether the tunneling dynamics succeeds in pushing $\langle\Delta\rangle\to 0$, or whether it is overwhelmed by the prior $p(\Delta)$, is directly determined by the choice of the measure on eternal inflation entering $p(\Delta)$, and decouples from the phase space problem of post-tunneling inflation.

\section*{Acknowledgments}
This work was supported by the Impuls und Vernetzungsfond of the
Helmholtz Association of German Research Centers under grant
HZ-NG-603, and German Science Foundation (DFG) within the
Collaborative Research Center 676 ``Particles, Strings and the Early
Universe''.

\bibliographystyle{kp}
\bibliography{tunneling_quartic}

\begingroup\raggedright\begin{thebibliography}{15}
\expandafter\ifx\csname natexlab\endcsname\relax\def\natexlab#1{#1}\fi

\bibitem[Huber and Konstandin(2008)]{Huber:2008hg}
S.~J. Huber and T.~Konstandin, ``{Gravitational Wave Production by Collisions:
  More Bubbles}'', {\em JCAP} {\bfseries 0809} (2008) 022,
  \href{http://xxx.lanl.gov/abs/0806.1828}{{\ttfamily arXiv:0806.1828}}.

\bibitem[Bousso and Polchinski(2000)]{Bousso:2000xa}
R.~Bousso and J.~Polchinski, ``{Quantization of four form fluxes and dynamical
  neutralization of the cosmological constant}'', {\em JHEP} {\bfseries 0006}
  (2000) 006,  \href{http://xxx.lanl.gov/abs/hep-th/0004134}{{\ttfamily
  arXiv:hep-th/0004134}}.

\bibitem[Susskind(2003)]{Susskind:2003kw}
L.~Susskind, ``{The Anthropic landscape of string theory}'',
  \href{http://xxx.lanl.gov/abs/hep-th/0302219}{{\ttfamily
  arXiv:hep-th/0302219}}.

\bibitem[Coleman(1977)]{Coleman:1977py}
S.~R. Coleman, ``{The Fate of the False Vacuum. 1. Semiclassical Theory}'',
  {\em Phys. Rev.} {\bfseries D15} (1977)
2929--2936.

\bibitem[Coleman and De~Luccia(1980)]{Coleman:1980aw}
S.~R. Coleman and F.~De~Luccia, ``{Gravitational Effects on and of Vacuum
  Decay}'', {\em Phys. Rev.} {\bfseries D21} (1980)
3305.

\bibitem[Hawking and Moss(1982)]{Hawking:1981fz}
S.~Hawking and I.~Moss, ``{Supercooled Phase Transitions in the Very Early
  Universe}'', {\em Phys.Lett.} {\bfseries B110} (1982) 35.

\bibitem[Callan and Coleman(1977)]{Callan:1977pt}
J.~Callan, Curtis~G. and S.~R. Coleman, ``{The Fate of the False Vacuum. 2.
  First Quantum Corrections}'', {\em Phys.Rev.} {\bfseries D16} (1977)
  1762--1768.

\bibitem[Adams(1993)]{Adams:1993zs}
F.~C. Adams, ``{General solutions for tunneling of scalar fields with quartic
  potentials}'', {\em Phys.Rev.} {\bfseries D48} (1993) 2800--2805,
  \href{http://xxx.lanl.gov/abs/hep-ph/9302321}{{\ttfamily
  arXiv:hep-ph/9302321}}.

\bibitem[Duncan and Jensen(1992)]{Duncan:1992ai}
M.~J. Duncan and L.~G. Jensen, ``{Exact tunneling solutions in scalar field
  theory}'', {\em Phys.Lett.} {\bfseries B291} (1992) 109--114.

\bibitem[Hamazaki et~al.(1996)Hamazaki, Sasaki, Tanaka, and
  Yamamoto]{Hamazaki:1995dy}
T.~Hamazaki, M.~Sasaki, T.~Tanaka, and K.~Yamamoto, ``{Selfexcitation of the
  tunneling scalar field in false vacuum decay}'', {\em Phys.Rev.} {\bfseries
  D53} (1996) 2045--2061,
  \href{http://xxx.lanl.gov/abs/gr-qc/9507006}{{\ttfamily
  arXiv:gr-qc/9507006}}.

\bibitem[Pastras(2011)]{Pastras:2011zr}
G.~Pastras, ``{Exact Tunneling Solutions in Minkowski Spacetime and a Candidate
  for Dark Energy}'',
 \href{http://xxx.lanl.gov/abs/1102.4567}{{\ttfamily arXiv:1102.4567}}.

\bibitem[Dutta et~al.(2011)Dutta, Vaudrevange, and Westphal]{Dutta:2011ej}
K.~Dutta, P.~M. Vaudrevange, and A.~Westphal, ``{An Exact Tunneling Solution in
  a Simple Realistic Landscape}'',
  \href{http://xxx.lanl.gov/abs/1102.4742}{{\ttfamily arXiv:1102.4742}}, *
  Temporary entry *.

\bibitem[Lee and Weinberg(1986)]{Lee:1985uv}
K.-M. Lee and E.~J. Weinberg, ``{TUNNELING WITHOUT BARRIERS}'', {\em
  Nucl.Phys.} {\bfseries B267} (1986) 181.

\bibitem[Dong and Harlow(2011)]{Dong:2011gx}
X.~Dong and D.~Harlow, ``{Analytic Coleman-de Luccia Geometries}'',
  \href{http://xxx.lanl.gov/abs/1109.0011}{{\ttfamily arXiv:1109.0011}}.

\bibitem[Dutta et~al.(2011)Dutta, Vaudrevange, and Westphal]{Dutta:2011fe}
K.~Dutta, P.~M. Vaudrevange, and A.~Westphal, ``{The Overshoot Problem in
  Inflation after Tunneling}'',
  \href{http://xxx.lanl.gov/abs/1109.5182}{{\ttfamily arXiv:1109.5182}}, *
  Temporary entry *.

\end{thebibliography}\endgroup
\end{document}